\documentclass[showpacs,amsmath,amssymb,pre,aps,superscriptaddress,twocolumn]{revtex4-1}
\usepackage[tight,TABTOPCAP]{subfigure}
\usepackage[usenames, dvipsnames]{color}
\usepackage{natbib}
\usepackage[utf8]{inputenc}
\usepackage{textcomp}
\usepackage[english]{babel}
\usepackage{amsmath}
\usepackage{amsfonts}
\usepackage{amssymb}
\usepackage{makeidx}
\usepackage{graphicx}
\usepackage{lipsum}
\usepackage{tikz}
\usepackage{float}
\usetikzlibrary{arrows}
\usetikzlibrary{decorations.pathreplacing,decorations.markings}
\usetikzlibrary{mindmap, trees}
\usetikzlibrary{shapes.geometric}
\usepackage{pgfplots}
\pgfplotsset{compat=1.16}
\usetikzlibrary{3d}
\usepackage[left=2cm,right=2cm,top=2cm,bottom=2cm]{geometry}
\usepackage{hyperref}
\usepackage{comment}

\begin{document}

\title{Density Fluctuations, Solvation Thermodynamics and Coexistence Curves in Grand Canonical Molecular Dynamics Simulations}
\author{Mauricio Sevilla}
\affiliation{Max Planck Institute for Polymer Research, Ackermannweg 10, 55128, Mainz, Germany}
\author{Luis A. Baptista}
\affiliation{Max Planck Institute for Polymer Research, Ackermannweg 10, 55128, Mainz, Germany}
\author{Kurt Kremer}
\email{kremer@mpip-mainz.mpg.de}
\affiliation{Max Planck Institute for Polymer Research, Ackermannweg 10, 55128, Mainz, Germany}
\author{Robinson Cortes-Huerto}
\email{corteshu@mpip-mainz.mpg.de}
\affiliation{Max Planck Institute for Polymer Research, Ackermannweg 10, 55128, Mainz, Germany}
\date{\today}
\begin{abstract}
Fluid transport across nanometric channels induced by electric, pressure and concentration gradients is ubiquitous in biological systems and fosters various applications. In this context, computer simulation setups with well-defined open-boundary equilibrium starting states are essential in understanding and assisting experimental studies. However, open-boundary computational methods are scarce and typically do not satisfy all the equilibrium conditions imposed by reality. Namely, in the absence of external gradients, 1) the system of interest (SoI) must be at thermodynamic and chemical equilibrium with an infinite reservoir of particles, 2) the fluctuations of the SoI in equilibrium should sample the grand canonical ensemble, 3) the local solvation thermodynamics, which is extremely sensitive to finite-size effects due to solvent depletion, should be correctly described. This point is particularly relevant for out-of-equilibrium systems. Finally, 4) the method should be robust enough to deal with phase transitions and coexistence conditions in the SoI. In this study, we demonstrate with prototypical liquid systems embedded into a reservoir of ideal gas particles that the adaptive resolution simulation (AdResS) method, coupled with particle insertion/deletion steps, satisfies all these requirements. Therefore, this AdResS setup is suitable for performing equilibrium and non-equilibrium simulations of open systems. 
\end{abstract}
\maketitle

\section{Introduction}
The study of fluid flow at the nanometric scale, i.e. nanofluidics~\cite{Emmerich_etal_NatRevMethPrim4_69_2024}, has recently gained considerable attention due to the prevalent role in biological systems~\cite{Yellen_Nature419_35_2002,Gouaux_Mackinnon_Science310_1461_2005} and a broad range of applications~\cite{Bell_etal_NanoLett12_512_2012}, ranging from energy harvesting~\cite{Feng_etal_Nature536_197_2016} to ionic computing~\cite{Robin_etal_Science379_161_2023}. Despite the highly developed experimental techniques, the interplay of confinement, non-equilibrium conditions, surface-specific interactions and quantum effects~\cite{Kavokine_etal_Nature602_84_2022,Yu_etal_NatNanotech18_898_2023} imposes significant challenges to the complete theoretical understanding of nanofluidics phenomena\cite{Schoch_etal_RevModPhys80_839_2008,Bocquet_Charlaix_ChemSocRev39_1073_2010,Kavokine_etal_AnnRevFluidMech53_377_2021}. In this context, open-boundary molecular dynamics (MD) approaches~\cite{Denniston_Robbins_PRL87_178302_2001,Denniston_Robbins_JChemPhys125_17166010_2006,Claudio2015,Lin_etal_PhysRevFluids1_17166010_2016,Hiroaki_etal_JChemPhys146_194702_2017,Liu_etal_PRL119_224502_2017,cortes2020} open the possibility to accompany, design and interpret experimental efforts.\\\\ 
Nevertheless, the non-equilibrium, open-boundary conditions present in experiments cannot be readily imposed on standard computational methods. Even the modest simulation of the corresponding equilibrium state must sample the grand canonical ensemble. The vast family of grand canonical Monte Carlo (MC)~\cite{Adams_MolPhys28_1241_1974,Adams_MolPhys29_307_1975,Mezei_MolPhys40_901_1980,Panagiotopoulos_IntJThermPhys10_447_1989,Shelley_Patey_JChemPhys100_8265_1994,Smit_MolPhys85_153_1995,Shelley_Patey_JChemPhys102_7656_1995,Shi_Maginn_JCTC3_1451_2007,Garberoglio_JChemPhys128_134109_2008,Soroush_etal_JChemPhys149_072318_2018,Fathizadeh_Elber_JChemPhys149_072325_2018} and hybrid MD/MC~\cite{Duane_etal_PhysLettB195_216_1987,Mehlig_etal_PRB45_679_1992,Boinepalli_JChemPhys119_12769_2003,Stern_JChemPhys126_164112_2007,Nilmeier_etal_PNAS108_E1009_2011,Chen_Roux_JChemPhys141_114107_2014,Ross_etal_JPhysChemB122_5466_2018,Belloni_JChemPhys152_021101_2019,Jeongmin_etal_JChemPhys159_144802_2023} techniques, which rely upon particle insertion steps that become cumbersome when increasing the system's density and molecular complexity, are not mean to impose gradients on open-boundaries. Thus, developing and testing open-boundary computational techniques that sample the grand canonical ensemble in equilibrium is paramount. To this aim, we identify four essential conditions that open-boundary MD methods aiming at investigating non-equilibrium phenomena should fulfil:\\\\
\textit{Constant chemical potential} - The molecular species in the system of interest (SoI) should be in thermal and chemical equilibrium with an infinite reservoir of particles. To the best of our knowledge, among the open-boundary MD methods available in the literature, only the adaptive resolution~\cite{DelleSiteChemPot,Heidari_Spatially2018,cortes2020,Baptista2021} and the constant $\mu$ methods~\cite{Claudio2015,karmakar_non-equilibrium_2023} partially satisfy this condition since the total size of the system limits the size of the reservoir under consideration. The next three requirements illustrate the relevance of the infinite size of the reservoir for open-boundary simulations.\\\\
\textit{Grand canonical ensemble} - The fluctuations of the number of particles, $\Delta^2 N /\langle N\rangle $, in the SoI should sample the grand canonical ensemble. More specifically, given a single-component SoI of volume $V_0\gg V_{\zeta}$, with $V_{\zeta}$ the volume define by the system's correlation length $\zeta$, embedded into an infinite reservoir of particles,  the quantity $\Delta^2 N /\langle N\rangle $ should satisfy the finite-size isothermal compressibility equation for every subvolume $V$ in $V_0$~\cite{Roman_etal_EurophysLett42_371_1998,Sevilla_CortesHuerto_JChemPhys156_044502_2022}

\begin{equation}
\label{eq:OZsize}
\begin{split}
&\Delta^2 N /\langle N\rangle=\chi_{T}(V;V_{0}) \\
&=  1 + \frac{\rho}{(2\pi)^3V} \int d \mathbf{k}\, \tilde{R}(\mathbf{k})\, \tilde{R}(-\mathbf{k})\, \tilde{h}^{\rm PBC}(\mathbf{k};V_0)\, .
\end{split}
\end{equation}

In this expression, $\Delta^2 N\equiv (\langle N^2 \rangle -\langle N \rangle^2)$ and $\langle N \rangle$ represents the average number of particles in the subvolume $V \le V_0$ such that the average density is $\rho=\langle N \rangle/V$. In the asymptotic limit, $|\mathbf{r}|>\zeta$, the radial distribution function of the open system, $g(\mathbf{r})$, is connected to the one computed in the closed system, $g(\mathbf{r};N_{\rm Tot})$, by the relation $g(\mathbf{r};N_{\rm Tot})=g(\mathbf{r}) - \rho k_{\rm B}T\kappa_T/N_{\rm Tot}$ with $\kappa_T$ being the bulk isothermal compressibility~\cite{Percus1961,Percus1961II,Salacuse-etal-PRE53-2382-1996,Salacuse-PhysA387-3073-2008,Villamaina-Trizac-EurJPhys35-035011-2014,Cortes2016,Cortes2018e,cortes2018m}. Hence, finite-size effects due to using different statistical ensembles can be easily identified by either considering a system with a finite total number of particles $N_{\rm Tot}$ (canonical) or including an infinite reservoir, i.e. $N_{\rm Tot}\to \infty$, (grand canonical). For simplicity, $h(\mathbf{r})=g(\mathbf{r})-1$ is defined, and ensemble finite-size effects are corrected when introducing explicitly the asymptotic limit of the radial distribution function. The integral on the right-hand side of the equation is evaluated in Fourier space so that $\tilde{h}$ is the Fourier transform of $h$. The step function $R(\mathbf{r})$ accounts for the finite integration subdomain, and $\tilde{R}$ corresponds to its Fourier transform.\\\\
An additional advantage of evaluating the integral in Fourier space is that explicit finite-size effects due to the use of periodic boundary conditions (PBC) can be trivially introduced via 
\begin{equation}\label{eq:hPBC}
\tilde{h}^{\rm PBC}(\mathbf{k};V_0)=\sum_{n_x,n_y,n_z} \exp(-\mathbf{k}\cdot\mathbf{s}_{n_x,n_y,n_z}) \tilde{h}(\mathbf{k})\, ,
\end{equation}
with $\mathbf{s}_{n_x,n_y,n_z}=(n_x\, L_{0x},n_y\, L_{0y},n_z\, L_{0x})$ a vector specifying the system's periodic images such that $n_{x,y,z}$ takes integer values and $V_0=L_{0x}\times L_{0y}\times L_{0z}$, where an orthorhombic simulation box is assumed. The choice $|n_x|\le 1$, $|n_y|\le 1$ and  $|n_z|\le 1$ is sufficient to compute Eq.~\eqref{eq:OZsize} accurately. To sum up, Eq.~\ref{eq:OZsize} is computed for a reference simulation and used as a benchmark. Fluctuations of the number of particles $\Delta^2 N/\langle N \rangle$ are computed for the SoI, and if the two results coincide for every $V$ in $V_0$, the SoI samples the grand canonical ensemble. Moreover, in the limit $V\to V_0$, 
$\Delta^2 N /\langle N \rangle$ should converge to $\rho k_{\rm{B}}T \kappa_T + \rm{constant}\times \rho/V_{0}^{1/3}$ in the grand canonical ensemble. By contrast, in the canonical ensemble $\Delta^2 N \to \rm{constant}\times \rho/V_{0}^{1/3}$~\cite{Cortes2016,Cortes2018e,cortes2018m}.\\\\
\textit{Solvation thermodynamics} - The simulation should be free of depletion effects that result from exhausting the number of particles in the reservoir~\cite{Mukherji-Kremer-Macromolecules46-9158-2013}. In practice, these effects can be monitored by evaluating Kirkwood-Buff integrals (KBI). Let us consider a binary mixture of $a$ and $b$ particles in a volume $V_0\gg V_{\zeta}$, embedded into an infinite reservoir. For every subvolume $V\le V_0$, the finite-size KBI, $G_{ij}(V;V_0)$ satisfies~\cite{Sevilla_CortesHuerto_JChemPhys156_044502_2022} 
\begin{equation}\label{eq:KBI_Fourier_PBC}
\begin{split}
&V\left(\frac{\left\langle N_{i} N_{j}\right\rangle-\left\langle N_{i}\right\rangle\left\langle N_{j}\right\rangle}{\left\langle N_{i}\right\rangle\left\langle N_{j}\right\rangle}-\frac{\delta_{i j}}{\left\langle N_{i}\right\rangle}\right)=
G_{i j}(V;V_0) \\
&\quad =\frac{1}{(2\pi)^3V} \int d \mathbf{k}\, \tilde{R}(\mathbf{k})\, \tilde{R}(-\mathbf{k})\, \tilde{h}^{\rm PBC}_{ij}(\mathbf{k};V_0)\, ,
\end{split}
\end{equation}
with $\delta_{ij}$ the Kronecker delta, $\langle N_{i}\rangle$ the average number of particles in the volume $V$ and the indices $i,j$ indicate the type of particle $a$,$b$. As for the single component case, the grand canonical ensemble can be characterised in this equation by taking the limit for the total number of particles $N_{\rm Tot}\to \infty$ in the two-component radial distribution function~\cite{ben-naim}.\\\\  
In practice, a benchmark simulation allows the evaluation of the integral in Eq.~\eqref{eq:KBI_Fourier_PBC}, which reduces to integrating the system's partial structure factors. Hence, a simulation samples the grand canonical ensemble if the fluctuations of the number of particles on the left-hand side of  Eq.~\eqref{eq:KBI_Fourier_PBC} follows the solution of the integral on the right-hand side of the same equation evaluated as a benchmark. Furthermore, both results should converge to  $G_{ij}+\rm{constant}/V_0^{1/3}$ with $G_{ij}$ being the KBI in the thermodynamic limit. Thus, the excess coordination number $N_{ij}$, i.e. the change in the number of $i$ particles resulting from removing a $j$ particle from the volume $V_0$,  should be $\approx \rho_i G_{ij}$. Instead, the corresponding limit for a closed system is $\approx -\delta_{ij}$.\\\\
\textit{Coexistence and critical conditions} - To investigate phase transformations in open molecular systems, it is crucial that the particle reservoir could be easily controlled to impose different thermodynamic conditions while maintaining thermal and chemical equilibrium with the SoI. We emphasise that near-critical conditions and second-order phase transitions have a limited description by any computational approach due to severe finite-size effects. Hence, let us explore the liquid-vapour coexistence region in the $T-\rho$ diagram of a single-component liquid. That means that the average density of the SoI should fluctuate between high-density (liquid) and low-density (vapour) values while keeping a constant chemical potential with the reservoir. From theoretical and practical perspectives, this point is crucial when simulating non-equilibrium conditions.\\
\begin{figure}[H]
	\centering    
	\includegraphics[width=0.45\textwidth]{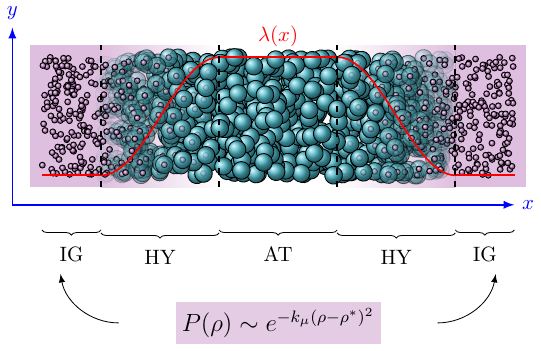}
	\caption{Schematic representation of the simulation setup. A slab geometry oriented along the $x$-axis with periodic boundary conditions applied in all directions. The switching field $\lambda(x)$ defines atomistic (AT), hybrid (HY) and ideal gas (IG) regions by taking values 1, $1>\lambda(x)>0$ and 0, respectively. A uniform density profile ensures the SoI has a constant chemical potential with the IG reservoir. The particle insertion/deletion algorithm is applied to the IG region, which is effectively in contact with an infinite particle reservoir. Periodic Boundary Conditions (PBC) are applied along $x$--, $y$-- and $z$--directions.}
	\label{fig:fig0}
\end{figure}
In this paper, we show that the Adaptive Resolution Simulation Method (AdResS) with particle insertion/deletion steps fulfils the conditions outlined above. To this aim, we simulate prototypical single-component and mixtures of Lennard-Jones systems embedded into infinite reservoirs of ideal gas particles that exhibit enough complexity (finite-size effects, non-trivial solvation behaviour, critical points, and a coexistence region) to test this AdResS setup extensively. 
\section{Model and Methods}
The AdResS Hamiltonian~\cite{Potestio_Hamiltonian2013,Potestio_Monte2013,Cortes2018} for a fluid composed of $N$ molecules containing $N_{a}$ atoms is
\begin{equation}
\label{eq:ham}
H_{[\lambda]}(r,p) = \mathcal{K} + V^{\rm intra} + \sum_{\alpha=1}^{N}\{ \lambda_{\alpha}V_{\alpha} + V^{\rm ext}(\lambda_{\alpha})\}\, ,
\end{equation}
with ($r,p$) positions and momenta and $\mathcal{K} = \sum_{i=1}^{N_{a}} \mathbf{p}_{i}^2/2m_{i}$ being the total kinetic energy of the system. Latin indices run over atoms, and Greek indices over molecules. The potential $V^{\rm intra} = \sum_{\alpha=1}^{N}\sum_{i\ne j \in \alpha}V^{\rm intra}(r_{ij})$ accounts for intra-molecular interactions, with $r_{ij}$ being the separation between atoms $i$ and $j$, which belong to the same molecule $\alpha$. Intermolecular interactions are included in the potential term $V_{\alpha} = \frac{1}{2}\sum_{\beta\neq \alpha}\sum_{i\neq j} V (|\mathbf{r}_{\alpha i}-\mathbf{r}_{\beta j}|)$ with $\mathbf{r}_{\alpha i}$ being the position of the atom $i$ in the molecule $\alpha$. The molecules' resolution is determined by the switching field $\lambda_{\alpha}\equiv \lambda(\mathbf{R}_{\alpha})$ with $\mathbf{R}_{\alpha}$ being the position of the centre of mass of the molecule $\alpha$. When $\lambda=0$ the Hamiltonian describes an homogeneous ideal gas system provided $V^{\rm ext}(0) = {\rm constant}$. In particular, we set $V^{\rm ext}(0) = 0$. For $0 < \lambda \le 1$, the Hamiltonian describes an inhomogeneous system, namely, an interacting system in the presence of an external field (Figure \ref{fig:fig0}).\\\\
By assuming that the SoI is in the grand canonical ensemble, namely, it is thermal and chemical equilibrium with an infinite reservoir of chemical potential $\mu^{\rm{id}}$, we have recently shown that the grand potential  can be written as a functional of the density \cite{Baptista2021}, i.e.
 \begin{equation}
 \begin{split}
 \Omega_{[\lambda]}[\rho^{[\lambda]}(\mathbf{r})] &= \\
 F_{[\lambda]}[\rho^{[\lambda]}(\mathbf{r})] &+ \int d\mathbf{r}\,  \rho^{[\lambda]}(\mathbf{r})\, (V^{\rm ext}(\lambda(\mathbf{r})) - \mu(\lambda(\mathbf{r}))) \, ,
 \end{split}
 \end{equation}
with the functional $F_{[\lambda]}[\rho^{[\lambda]}]$ being the intrinsic Helmholtz free energy corresponding to the Hamiltonian (\ref{eq:ham}), independent of the external potential. The condition that the SoI, i.e. the AT region in Fig.~\ref{fig:fig0}, is in thermal and chemical equilibrium with the reservoir can be imposed explicitly as $F_{[\lambda]}[\rho^{[\lambda]}]=F^{\rm{id}}[\rho(\mathbf{r})]$ with $F^{\rm{id}}$ the Helmholtz free energy of the ideal gas.  With this condition, the minimisation of the functional $\Omega_{[\lambda]}[\rho^{[\lambda]}(\mathbf{r})]$ implies 
\begin{equation}
\rho(\mathbf{r}) = \rho^{\rm{id}}\exp{(-\beta\{ V^{\rm ext}(\lambda(\mathbf{r})) - \mu^{\rm exc}(\lambda(\mathbf{r})) \})}\, ,
\end{equation}
with $\rho^{\rm{id}}$ being the density of the ideal gas. A useful condition, that guarantees that  $\rho(\mathbf{r}) = \rho^{\rm{id}}$ constant, is given by  $V^{\rm ext}(\lambda(\mathbf{r})) = \mu^{\rm exc}(\lambda(\mathbf{r}))$. Conversely, by classical density functional theory, if a thermodynamic force is applied to the molecules instantaneously present in the HY region such that the average density is constant across the simulation box, minus the integral of the force is equal to the excess chemical potential of the SoI with respect to the ideal gas at the same density and temperature. These results show that the AdResS method satisfies the first requirement concerning a constant chemical potential between the SoI and the reservoir. Moreover, unlike standard grand canonical Monte Carlo, the AdResS method does not require the system's chemical potential $\mu$ as an input parameter. This is an advantageous feature since $\mu$ is typically a difficult quantity to compute.\\\\
The infinite reservoir conditions could be approximated by increasing the size of the IG region as much as possible. However, it is exact and computationally more efficient to introduce Metropolis Monte Carlo particle insertion/deletion steps in the ideal gas (AdResS) such that the probability that the present density $\rho$ changes by a quantity $\delta\rho$, is given by \cite{cortes2020}
\begin{equation}
\text{acc}(\rho \to \rho\pm \delta\rho)=\text{min}[1,\text{exp}(-k_{\mu}\delta\rho(\delta\rho \pm 2(\rho-\rho^{\rm{id}})))]\, ,
\end{equation}
with $k_{\mu}$ a free parameter related to the width of the distribution of possible values of $\rho$ and $\rho^{\rm id}$ the reference density enforced in the IG region. \\\\
The remainder of the paper is devoted to using the AdResS approach to investigate the three requirements described above to simulate open systems with a well-defined equilibrium state.\\
\section{Results}
\subsection{Grand canonical ensemble} 
\begin{figure}[H]
	\centering
	\includegraphics[width=0.48\textwidth]{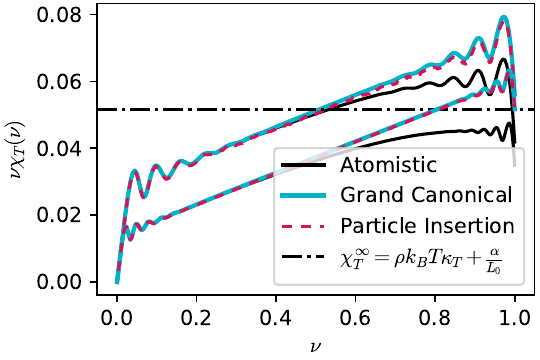}
	\caption{Fluctuations of the number of particles for two different simulation boxes defined by the vectors ($15\sigma$,$15\sigma$,$50\sigma$) and  ($30\sigma$,$30\sigma$,$80\sigma$). The solid blue curves correspond to the solution of the grand canonical finite-size integral in Eq.~\eqref{eq:OZsize} for the small (upper) and the large (lower) samples. The solid black curves are the corresponding solution in the canonical ensemble. The horizontal dashed-dotted black line represents the thermodynamic limit $\rho k_{\rm{B}}T \kappa_T$. The fluctuations of the number of particles in the AT region obtained within the AdResS framework are presented as dashed red curves. } 
	\label{fig:fig1}
\end{figure}

To prove that the simulation setup samples the grand canonical ensemble, it is sufficient to show that the fluctuations of the number of particles in the SoI follow Eq.~\ref{eq:OZsize}. We compute the benchmark integral in Eq.~\eqref{eq:OZsize} for a liquid system whose potential energy is described by a 12-6, truncated and shifted, Lennard-Jones (LJ) potential with cutoff radius $r_{\rm c}/\sigma = 2.5$. The parameters $\epsilon$, $\sigma$ and $m$ define the units of energy, length and mass, respectively. The results are expressed in LJ units with time $\tau = \sigma(m/\epsilon)^{1/2}$, temperature $\epsilon/k_{\rm B}$ and pressure $\epsilon/\sigma^3$. A cubic box of linear size $30\sigma$ is considered with density fixed at $\rho\sigma^3=0.864$. The system is equilibrated at $k_{\rm B}T/\epsilon = 2$, enforced by a Langevin thermostat with damping coefficient $\gamma(\sigma(m/\epsilon)^{1/2})=1$ for $15\times 10^{7}$ steps using a time step of $\delta t / (\sigma(m/\epsilon)^{1/2}) = 0.01$. Production runs span $15\times 10^{7}$ MD steps. All simulations were performed with an in-house version of the LAMMPS~\cite{LAMMPS} simulation package. \\\\ 
By calculating the structure factor, it is thus possible to evaluate the integral on the right-hand side of Eq.~\eqref{eq:OZsize}, correct the radial distribution function for ensemble effects and use the PBC term to extrapolate the result corresponding to the simulation box geometry of Fig.~\ref{fig:fig0} with the volume $V_0$ defined as the volume of the AT region. We consider two orthorombic simulation boxes defined by the vectors ($15\sigma$,$15\sigma$,$50\sigma$) and ($30\sigma$,$30\sigma$,$85\sigma$) and with the AT region of dimensions given by ($15\sigma$,$15\sigma$,$15\sigma$) and ($30\sigma$,$30\sigma$,$30\sigma$), respectively. The results of the integration for the two simulation boxes as a function of the ratio $\nu = (V/V_0)^{1/3}$ are presented as the solid blue curves in Fig.~\ref{fig:fig1}. The solid black curves represent the canonical ensemble cases for the same geometries, which are shown for comparison. The important difference between the two curves corresponds to the region $\nu\approx 1$ or $V\to V_0$, where the two limiting cases discussed in the introduction are apparent.\\\\ 
With this benchmark at hand, we compute the fluctuations of the number of particles in the AT region within the AdResS framework described above. Starting from fully atomistic samples, hybrid (HY) and ideal gas (IG) regions are defined with linear sizes $15\sigma$ and $\sigma$, respectively. The HY region is uniformly discretized in slabs of size $0.5\sigma$. To equilibrate the samples, the thermodynamic and drift forces needed to flatten the density profile across the simulation are updated every $10^4$ steps during $7\times10^7$ steps. After $4\times10^7$ steps, the density profile is considered converged, and the integral of the final thermodynamic and drift forces gives the excess chemical potential of the LJ liquid at these conditions, $\mu_{\rm exc}/\epsilon=8.866\pm0.0613$. At this stage, the particle insertion/deletion algorithm is applied every 1000 steps, and the simulation runs for further $10^7$ steps for production results.\\\\  
The result of computing $\Delta^2 N\equiv (\langle N^2 \rangle -\langle N \rangle^2)$ in the AT region for subvolumes $V\le V_0$ is presented as the dashed red curve in Fig.~\ref{fig:fig1} for the two simulation boxes considered.  The AdResS simulation reproduces the benchmark solid blue curve, thus demonstrating that the AT region samples the grand canonical ensemble.  As expected from the considerations in the introduction, the linear regime at intermediate values of $\nu$ converges to $\rho k_{\rm{B}}T \kappa_T$, represented by the horizontal dashed-dotted black line, plus a contribution $\propto 1/V_{0}^{1/3}$, which shifts down the curve for the larger system, approaching the thermodynamic limit value.
\subsection{Solvation thermodynamics}%
To investigate solvation thermodynamics, we focus on investigating a binary mixture ($a,b$) of a fluid whose potential energy is given by the purely repulsive truncated and shifted 12-6 LJ potential with cutoff radius $2^{1/6}\sigma$. The potential parameters are $\sigma_{aa} = \sigma_{bb}=\sigma_{ab} = \sigma$ and $\epsilon_{aa} = 1.2\epsilon$, $\epsilon_{bb} = 1.0\epsilon$ and $\epsilon_{ab} = (\epsilon_{aa}+\epsilon_{bb})/2$. As for the single-component case, the results are expressed in LJ units with energy $\epsilon$, length $\sigma$, and mass $m_a=m_b=m$. A constant temperature of $k_{\rm B}T/\epsilon = 1.2$ is enforced by the Langevin thermostat, as before.\\\\ 
The benchmark has been computed for a system with a total number of atoms $N_{\rm Tot} = 24000$, and the range of mole fractions of $a$-molecules $x_a = 0.2,\cdots,0.8$. The pressure is fixed at $P\sigma^3/\epsilon = 9.8$. After equilibration by alternating NVT-NPT runs, the final NVT production run corresponds to a cubic simulation box with volume $V_0$ being the average volume in the last NPT run such that pressure does not deviate significantly from the target value. With the resulting partial structure factors, the integral on the right-hand side of Eq.~\eqref{eq:KBI_Fourier_PBC} can be evaluated for arbitrary geometries of the simulation box, giving the behaviour of the local solvation environment for every subdomain $V\le V_0$. Furthermore, the thermodynamic limit value of the KBI, $G_{ij}$ is obtained.\\\\    
\begin{figure}[h]
	\centering
	\includegraphics[width=0.48\textwidth]{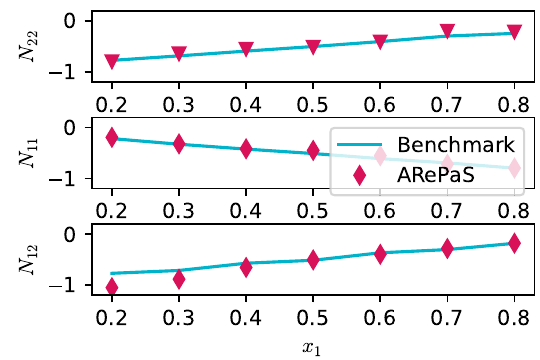}
	\caption{Excess coordination numbers $N_{ij}=\rho_i G_{ij}$ as a function of the concentration of $a$-molecules, $x_a$. The solid blue curves give the benchmark result. The red diamonds correspond to the result obtained by extrapolating the fluctuations of the number of particles in the AdResS simulation to evaluate $G_{ij}$.}
	\label{fig:fig2}
\end{figure}
We compute the fluctuations of the number of particles in the AT region using the AdResS setup. Orthorhombic fully atomistic simulation boxes were equilibrated using the NPT-NVT cycles described above. Using the final sizes for these systems, the benchmark result of computing the integral in Eq.~\eqref{eq:ham} is obtained. AT and HY regions of linear sizes $20\sigma$ and $10\sigma$, respectively, were defined. The remaining simulation parameters were taken from the previous single-component LJ liquid computations. The simulation converges when the values of the excess chemical potential for the $a,b$-molecules reproduce within the error bars the values reported in the literature ($\mu_{\rm exc}^{a,b}/\epsilon=(15.627\pm0.124, 15.374 \pm 0.122)$ for $x_1=0.5$, for example). Following this step, the particle insertion/deletion algorithm is applied every 1000 steps, and the production run corresponds to further $2\times10^8$ steps. \\\\
In this case, it was also verified that the fluctuations of the number of particles as computed by the left-hand side of Eq.~\eqref{eq:KBI_Fourier_PBC} reproduce the benchmark result obtained by evaluating the right-hand side of the same equation (Result not shown), indicating that the AdResS simulation reproduces the local and long-range solvation behaviour expected in the grand canonical ensemble. From the linear regime at intermediate values of $\nu$, the value of $G_{ij}$ is extrapolated. We present (Fig.~\ref{fig:fig2}) the excess coordination number $N_{ij}$ as calculated from the benchmark result (solid blue) and the AdResS (red diamonds) simulation. It is also apparent in this case that the two results agree well, indicating that the AdResS framework is robust enough to describe the system's solvation thermodynamics accurately. 
\subsection{Coexisting conditions}
\begin{figure}[h]
	\centering
	\includegraphics[width=0.48\textwidth]{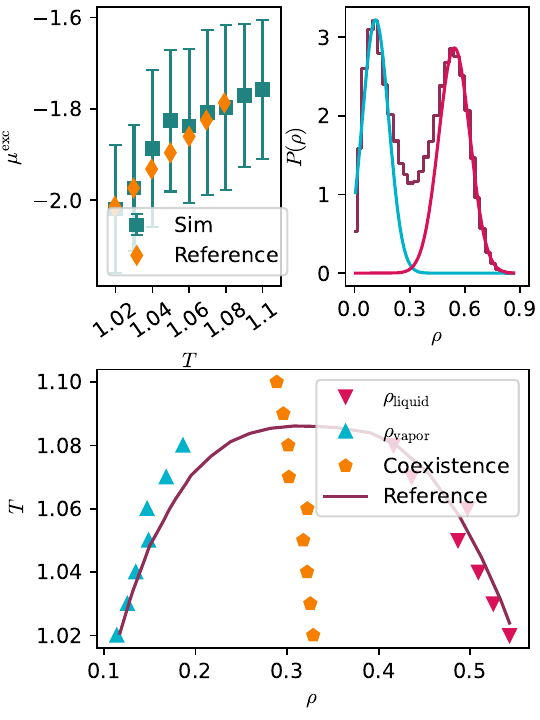}
	\caption{Critical point and coexistence conditions for the truncated and shifted LJ liquid. (a) Excess chemical potential $\mu_{\rm exc}$ as a function of temperature: our simulation data (red diamonds) and reference from Ref.~\cite{Wilding_PhysRevE_52_602_1995,Wilding_AmJPhys69_1147_2001}. (b) Probability distributions for the density of the coexisting liquid (red) and vapor (blue) systems at $k_{\rm B}T/\epsilon=1.02$. (c) Peak densities for the coexistent liquid (red) and vapor (blue) systems. The orange circles indicate the coexistence diameter. These data compare well with the reference from Ref.~\cite{thol_etal_IntJThermophys36_25_2015}.}
	\label{fig:fig3}
\end{figure}
Finally, we investigate the critical point transition and liquid-vapor coexistence conditions for the LJ system discussed in the first part of the manuscript. To this aim, we quenched the system at the critical density $\rho_{\rm c} \sigma^3 = 0.319$ in the range of temperatures $k_{\rm B}T/\epsilon=1.02,\cdots,1.10$, which includes the critical temperature $k_{\rm B} T_{\rm c}/\epsilon = 1.086$~\cite{thol_etal_IntJThermophys36_25_2015}.  \\\\
Results are presented in Fig.~\ref{fig:fig3}. First, the excess chemical potential as a function temperature across the coexistence line is calculated and compared with the values obtained with grand canonical Monte Carlo in Ref.~\cite{Wilding_PhysRevE_52_602_1995,Wilding_AmJPhys69_1147_2001}. Even though the error bars in our calculation are close to $5\%$, the overall trend displayed by the reference data is well reproduced by the AdResS simulation. Here, we emphasize that, in contrast to the grand canonical Monte Carlo method, the chemical potential is an output of the AdResS method, resulting from imposing an initial reference density.\\\\
We have computed density fluctuations in the AT region (See Fig.~\ref{fig:fig3}(b)). Below the critical point, the probability distribution shows the well-known bimodal distribution obtained with grand canonical Monte Carlo simulations, showing the system's tendency to fluctuate between liquid and vapor states. For the temperature range considered, we identified the maxima of these distributions and plotted the result as a function of the system's average density. The result is presented in Fig.~\ref{fig:fig3} (c) and compared with the grand canonical Monte Carlo results in Ref.~\cite{thol_etal_IntJThermophys36_25_2015}. It can be appreciated that the AdResS results reproduce the reference data reasonably well. 
\section{Conclusions}
Confinement, surface-specific interactions, and quantum effects make nanofluidics an ideal framework to challenge our understanding of non-equilibrium statistical mechanics. Computational methods are crucial to investigating these systems, aiming to interpret and drive experimental studies.
In this context, we argue that any open-boundary molecular dynamics simulation under the effect of external electric, density or concentration gradients should reduce to a well-defined equilibrium state when the external perturbation is not present. This state corresponds to the system of interest (SoI) being in chemical and thermal equilibrium with an infinite reservoir of particles. Hence, in equilibrium, the simulation setup must sample the grand canonical ensemble.\\\\ 
Specifically, for every subvolume $V$ within the volume $V_0$ of the SoI, the fluctuations of the number of particles should reproduce the corresponding finite-size integral equations evaluated under grand canonical conditions. Moreover, in the case of multicomponent systems, the solvation structure for every $V\le V_0$ should correspond to the finite-size Kirkwood-Buff integrals in the grand canonical ensemble. This condition ensures that the SoI is free from artefacts due to the depletion of particles in the reservoir. Finally, the simulation method should be robust enough to investigate the complexities of the phase diagram, such as near-critical conditions and coexistence curves.\\\\
We have simulated Lennard-Jones (LJ) liquids and mixtures in thermal and chemical equilibrium with an infinite reservoir of ideal gas particles using the AdResS method with particle insertion/deletion steps. In this setup, we have demonstrated that the fluctuations of the number of particles of the LJ systems sample the grand canonical ensemble for every subvolume $V\le V_0$. Furthermore, we have investigated the near critical point and liquid-vapor coexistence conditions, and the results presented compare well with existing data. Therefore, we conclude that this AdResS framework is suitable for investigating equilibrium open molecular liquids and mixtures. The simplicity of the ideal gas reservoir becomes a tremendous advantage when imposing different conditions, such as pressure and concentration gradients and external potentials, that can be applied simultaneously on the SoI. These studies will be the subject of forthcoming papers.  

\begin{acknowledgments}
L.A.B., K.K. and R.C.-H. thankfully acknowledge funding from SFB-TRR146 of the German Research Foundation (DFG) -- Project number 233630050. K.K. and R.C.-H. thank Maziar Heidari and Raffaello Potestio for a fruitful collaboration that constitutes the basis of the present work. R.C.-H. also thank Debashish Mukherji, Nancy C. Forero-Martinez, Kostas Daoulas and Pietro Ballone for many stimulating discussions. Simulations have been performed on the THINC cluster at the Max Planck Institute for Polymer Research. 
\end{acknowledgments}
\vspace{0.5cm}
\textbf{Data Availability Statement}\\
Data will be made available upon reasonable request.\\\\

\bibliographystyle{unsrt}

%
\end{document}